\documentclass{IEEEcsmag}

\usepackage{multirow}
\usepackage{multicol}
\usepackage{listings}
\usepackage{xcolor}
\usepackage{cite}   %
\definecolor{lightblue}{rgb}{0.81, 0.83, 0.92} %
\usepackage[backgroundcolor=white]{todonotes}
\usepackage[colorlinks,urlcolor=blue,linkcolor=blue,citecolor=blue]{hyperref}

\jvol{XX}
\jnum{XX}
\paper{8}
\jmonth{March}
\jname{Computing in Science and Engineering}
\pubyear{2022}

\let\templatesubsection\subsection
\renewcommand{\subsection}[1]{\templatesubsection{\textbf{#1}}}

\begin{document}
\thispagestyle{empty}
\pagenumbering{gobble}
\textbf{This paper has been accepted for publication as
"The State of Fortran," in Computing in Science \& Engineering. doi:10.1109/MCSE.2022.3159862}\\

\paragraph{IEEE Copyright Notice}
© 2022 IEEE. Personal use of this material is permitted.
Permission from IEEE must be obtained for all
other uses, in any current or future media, including reprinting/
republishing this material for advertising
or promotional purposes, creating new collective works, for
resale or redistribution to servers or lists, or
reuse of any copyrighted component of this work in other works.\\

\begin{lstlisting}
@article{stateOfFortran2022,
author={Kedward, Laurence and Aradi, Balint and Certik, Ondrej and
Curcic, Milan and Ehlert, Sebastian and Engel, Philipp and Goswami,
Rohit and Hirsch, Michael and Lozada-Blanco, Asdrubal and Magnin,
Vincent and Markus, Arjen and Pagone, Emanuele and Pribec,
Ivan and Richardson, Brad and Snyder, Harris,
and Urban, John and Vandenplas, Jeremie},
journal={Computing in Science \& Engineering},
title={The State of Fortran},
year={2022},
doi={10.1109/MCSE.2022.3159862}}
\end{lstlisting}

\clearpage\pagenumbering{arabic}
\sptitle{Feature Article}
\editor{Editor: Konrad Hinsen, konrad.hinsen@cnrs.fr}

\title{\vspace{-0.7em} The State of Fortran \\ \vspace{-1.5em}}

\author{{Laurence} Kedward}
\affil{\scriptsize Department of Aerospace Engineering, University of Bristol, Bristol, UK\\
Corresponding author: laurence.kedward@bristol.ac.uk}

\author{B\'alint Aradi}
\affil{\scriptsize Bremen Center for Computational Materials Science, University of Bremen, Germany}

\author{Ond\v{r}ej \v{C}ert\'{i}k}
\affil{\scriptsize Los Alamos National Laboratory}

\author{Milan Curcic}
\affil{\scriptsize University of Miami, Miami, FL, USA}

\author{Sebastian Ehlert}
\affil{\scriptsize Mulliken Center for Theoretical Chemistry, Institut für Physikalische und Theoretische Chemie, Universit\"{a}t Bonn, Germany}

\author{Philipp Engel}
\affil{\scriptsize  Institut für Geod\"{a}sie und Geoinformationstechnik, Technische Universit\"{a}t Berlin, Germany}

\author{Rohit Goswami}
\affil{\scriptsize Quansight Austin, TX, USA and Science Institute, University of Iceland}

\author{Michael Hirsch}
\affil{\scriptsize Center for Space Physics, Boston University, Boston, Massachusetts, USA}

\author{Asdrubal Lozada-Blanco}
\affil{\scriptsize  S\~{a}o Carlos Institute of Physics, University of S\~{a}o Paulo, S\~{a}o Carlos, SP, Brazil}

\author{Vincent Magnin}
\affil{\scriptsize  Univ. Lille, CNRS, Centrale Lille, Univ. Polytechnique Hauts-de-France, IEMN, Lille, France}

\author{Arjen Markus}
\affil{\scriptsize Deltares Research Institute, The Netherlands}

\author{Emanuele Pagone}
\affil{\scriptsize Cranfield University, Sustainable Manufacturing Systems Centre, School of Aerospace Transport and Manufacturing, Cranfield, UK}

\author{Ivan Pribec}
\affil{\scriptsize Chair of Brewing and Beverage Technology, Technical University of Munich, Germany}

\author{Brad Richardson}
\affil{\scriptsize Archaeologic, Inc., CA, USA}

\author{Harris Snyder}
\affil{\scriptsize Structura Biotechnology Inc., Toronto, Ontario, Canada}

\author{John Urban}
\affil{\scriptsize HPC Consultant, USA}

\author{J\'{e}r\'{e}mie Vandenplas}
\affil{\scriptsize Animal Breeding and Genomics, Wageningen UR, P.O. 338, 6700 AH, Wageningen, The Netherlands \\ \vspace{-5em}}

\markboth{Scientific Programming}{Paper title tbd}

\begin{abstract}
\small 
A community of developers has formed to modernize the Fortran ecosystem.
In this article, we describe the high-level features of Fortran that continue to make it a good choice for scientists and engineers in the 21st century.
Ongoing efforts include the development of a Fortran standard library and package manager, the fostering of a friendly and welcoming online community, improved compiler support, and language feature development.
The lessons learned are common across contemporary programming languages and help reduce the learning curve and increase adoption of Fortran.
\end{abstract}

\maketitle

\chapterinitial{Fortran} is a high-level programming language primarily used to solve scientific and engineering problems.
It has been under active development since its inception under John Backus at IBM in 1954 to the present day.
The initial goal was to ease the translation of mathematical formulas to optimized machine code instructions, a concept now known as compilation.
The intuitive abstraction of mathematical procedures enabled rapid development of numerical solutions to scientific problems, at a time when most programs were still hand-coded in assembly language.
Following the release of its first implementation in 1957, the language was
adopted by the scientific and engineering communities for writing numerical programs.
As a result, the language was quickly ported to several computer architectures such that Fortran is accepted as being the first cross-platform programming language.

The ISO Fortran Standard and its maintenance of backwards compatibility provide guarantees for language stability and code longevity.
Consequently, there is a mature and well-established ecosystem of Fortran compilers and libraries\textemdash
what is lesser known is how much the language has evolved since its beginnings.
In this article, we first review the high-level features that continue to make Fortran a powerful and effective tool for high-performance, scientific computing.
Second, we present recent progress and efforts towards uniting the Fortran community and improving the tooling and resources available to Fortran programmers.
Finally, we provide an outlook on the future of Fortran and its burgeoning ecosystem.

\section{WHY FORTRAN?}

\begin{figure*}[t]
	\centering
	\includegraphics[width=0.95\textwidth]{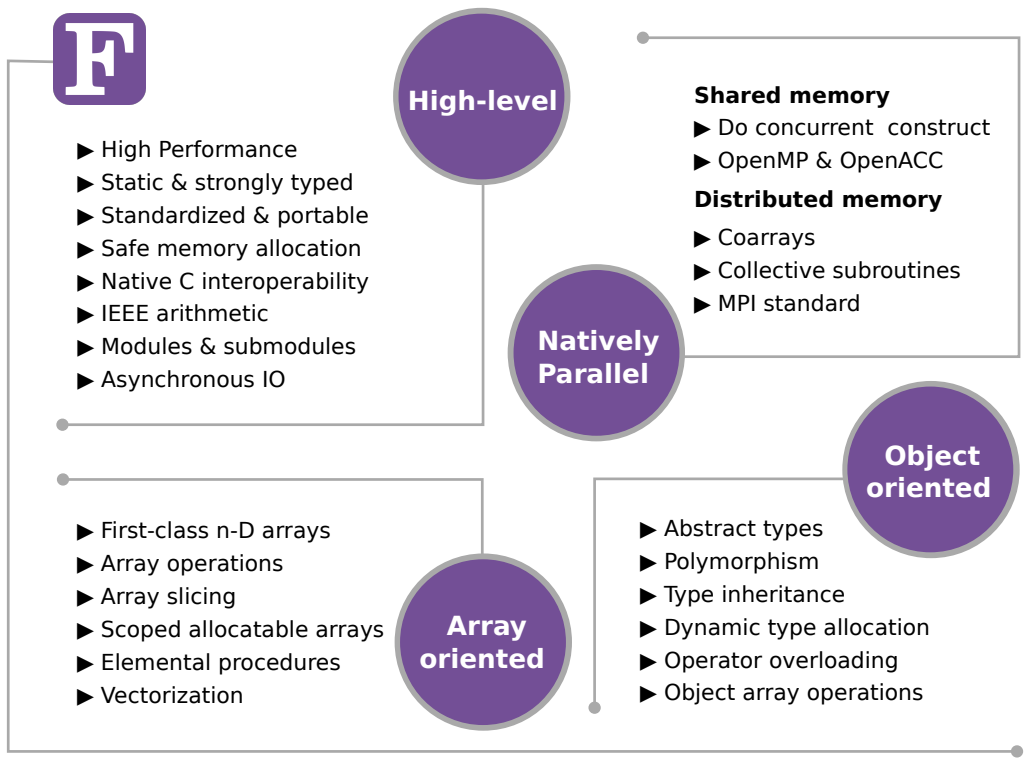}
	\caption{High-level language features of Fortran}
	\label{fig:fortran_features}
\end{figure*}

Fortran is perceived in some software development circles as archaic,
lacking the features and conveniences of newer languages, and characterized
by an obtuse syntax.
However, such considerations typically stem from a lack of familiarity with
Fortran standards later than Fortran 77 \cite{why_fortran_decyk07}.
Similarly, later updates to the Fortran language competed with newer alternatives such as C, MATLAB and Python that built on the high-level concepts first introduced in Fortran.

The Fortran language has seen many revisions \cite{Metcalf2011,Reid2007,Reid2014,Reid2018}, with the latest international standard being Fortran 2018.
Though a large body of Fortran code written in the 1970s to 1980s remains in use to this day, the language and its paradigms have evolved significantly.
In the remainder of this section we present the key features that make Fortran an effective tool for high-performance numerical computing.

\subsection{High-Level Language Features}

Key to the early popularity and continued use of Fortran are the high-level abstractions provided by the language (Figure \ref{fig:fortran_features}).
Intrinsic support for multidimensional arrays in combination with safe dynamic allocation and expressive array slicing, makes for a readable and easy-to-learn language suitable for high-performance numerical applications.
Strong, static typing helps avoid common programming pitfalls via meaningful compile-time errors without sacrificing runtime performance.
Similarly, modules and submodules allow for code organisation and reuse with automatic checking of routine signatures and types.
Fortran \texttt{allocatable} arrays enable dynamic allocation without memory leaks and are guaranteed to be non-aliasing.
As a result, efficient machine code can be generated by the compiler handling \texttt{allocatable} arrays, without requiring additional annotations such as the C \texttt{restrict} keyword.
Moreover, the array-oriented design of the language means that operator overloading with object arrays can be optimized by the compiler without requiring complex expression templates as in C++.
Additional language keywords that aid compiler optimization and ensure compile-time correctness include: the \texttt{intent} attribute for indicating whether procedure arguments are inputs, outputs, or bidirectional;
the \texttt{pure} attribute indicates procedures that have no external side effects; and
the \texttt{elemental} attribute maps a procedure to each element of an N-dimensional array or scalar, useful for rank polymorphism and akin to the map operation in functional programming languages.

Fortran natively supports arrays up to rank 15, complex numbers, object-oriented programming, and syntax for expressing shared-memory and distributed-memory parallelism.
The Fortran native parallelism offers compiler-enforceable safeguards against many pitfalls of parallel programming.
Shared-memory parallelism for `do` loops, which has been supported through the OpenMP standard for many years, can now be expressed with the new \texttt{do concurrent} syntax to inform the compiler that loop iterations may be executed in any order using threading, SIMD (single instruction, multiple data), or any other parallel implementation available to the compiler.
For example, the freely available NVIDIA \textit{nvfortran} compiler can automatically offload standard \texttt{do concurrent} loops for execution to NVIDIA GPUs.
Listing \ref{lst:doconcurrent} shows a two-dimensional loop using \texttt{do concurrent}.

\begin{figure}
\begin{lstlisting}[language=Fortran,caption=Do concurrent example,label={lst:doconcurrent},morekeywords={concurrent}]
do concurrent (i = 1:n, j = 1:m)
  u(i,j) = a * v(j,i)
end do
\end{lstlisting}
\end{figure}

Distributed-memory parallelism has typically been implemented in Fortran using the MPI standard.
Fortran 2008 introduced intrinsic abstractions for distributed-memory parallelism in the form of \textit{coarrays}.
Parallel Fortran programs follow the SPMD (single program, multiple data) paradigm, where
multiple copies of the same executable are started as different processes, with
each process working on a different piece of the problem.  These
processes (\textit{images}, in Fortran terminology) have independent
address spaces but can exchange data using coarrays or collective subroutines, introduced in Fortran 2018.  If a variable is
declared as a coarray, an image can
read and write the value of that variable on other images with a simple
syntax that is similar to array indexing (an image number is used instead of an
array index).
Synchronizing images is as simple as issuing a \texttt{sync all} statement.
More fine-grained control is possible with subsets of workers known as \textit{teams},
and event-based scheduling with the \textit{events}.
Coarrays, collectives, teams, and events
abstract away much of the verbosity required by MPI thereby helping programmers
to focus on their application rather than the low-level details of
interprocess data exchange.  Several coarray implementations are
based on MPI, allowing Fortran programmers to leverage the excellent hardware
interconnects supported by MPI libraries. Reid \textit{et al.}
summarize the evolution of Fortran's parallel features~\cite{Reid2020history}.

The \texttt{iso\_c\_binding} module is an essential feature introduced in
Fortran 2003 and has been further extended in subsequent versions. It provides
standardized interoperability with the C programming language to access
procedures, structures, and global variables to and from Fortran.  Nearly any
programming language that has a C interface such as Python or C++
can interact with compiled
Fortran code in a platform-agnostic manner.  Foreign functions are bound through
interface declarations that often can be generated automatically and are called
like any other native Fortran procedure.  The resulting application is linked
against the corresponding C library or object file.  Consequently, the
programmer can reuse source code in a mixed-language environment or can
outsource tasks like connecting to hardware via OEM drivers written in C.  A
broad range of interface libraries is already available, including common
use-cases: system libraries (libc, POSIX); toolkits (GTK, Motif); abstraction
layers (OpenGL, SDL 2.0, Curses); scripting languages (Lua, Python, MATLAB); and
networks (cURL, FastCGI).  The \texttt{iso\_c\_binding} does not cover all C99
features\textemdash{}for instance, variadic functions which have no equivalent in
Fortran\textemdash{}however the binding continues to be under development for
future Fortran revisions.

Additionally, Fortran procedures can be exposed to C or any other programming
language with foreign function interface, such as Python, Lua, or Rust.
Interoperable functions and subroutines are indicated by the \texttt{bind(c)}
attribute and the standardized \texttt{ISO\_Fortran\_binding.h} header allows calling Fortran
procedures from C or other languages using a C structure
to interface with Fortran array descriptors.

\subsection{Stability and Reliability}
As an actively used language with a long history, Fortran and its associated compilers have long maintained excellent backward compatibility and language stability.
This is important for many well-established and large Fortran projects that are critical for numerically intensive tasks.
For such large projects, a fast-developing language with breaking API changes would incur significant costs to the maintainers in person-hours of redevelopment and revalidation time.
By contrast, Fortran offers inherent guarantees of code longevity and indeed many projects in use today are largely unchanged from their original release despite being written under Fortran standards dating from decades ago.

Importantly, anyone writing Fortran software today can have confidence not only in the longevity of their code but also in that of any Fortran libraries and applications they use.
This inherent future-proofing of Fortran code therefore protects against so-called \textit{code rot} or \textit{software collapse} \cite{Hinsen2019} whereby code becomes unusable due to breaking changes in its dependencies or its implementation language.

\subsection{Mature}

A key accomplishment of Fortran 90 and 95 was to incorporate dynamic arrays and \textit{de facto} procedures and syntax widely adopted by compilers of the time.
Along with the C interoperability mentioned previously, Fortran 2003 brought improved operations for interacting with the environment, such as reading environment variables, reading command-line arguments, and running external programs.
Remaining gaps in systems programming such as filesystem operations are
being addressed in the growing Fortran standard library described in the following section.
While some projects have recently moved from Fortran to C++ or other languages, the underpinnings of numerical computing remain implemented in Fortran.
Often a user may not be aware they are using Fortran via another language such as the Python NumPy or SciPy libraries which are widely used among the scientific community.

There are several mature Fortran compilers, many of which support the latest Fortran standards.
The OpenMP and OpenACC standards provide directives for parallel execution in a platform-agnostic manner.
NVIDIA promotes CUDA Fortran and OpenACC in the \texttt{nvfortran} compiler for offloading of parallel constructs to the GPU.
The MPI-3 standard continues to support modern Fortran syntax with polymorphic interfaces scalable from multi-core embedded systems to the largest supercomputers.

New books continue to be written on modern Fortran, major funding agencies across the globe continue to fund upgrades and new projects in Fortran, and graduate students in many STEM disciplines continue to learn Fortran.
The Fortran community, as discussed in the next section, has renewed vigor and is rapidly growing the Fortran ecosystem using best practices of modern software development.

It is for these reasons\textemdash
performance, ease-of-use, productivity, portability, stability and longevity\textemdash
that we advocate for Fortran.
In the remainder of the paper, we describe the shortcomings of Fortran and its ecosystem, and present the ongoing community efforts to address them.

\section{FORTRAN-LANG}
\subsection{A New Community for Fortran Users}

Despite recent revisions to the language and continuing to be a large share of the software executed on major
High Performance Computing (HPC) systems, Fortran's ecosystem has stagnated
across multiple fronts.
First, the lack of a standard library, a common resource in modern programming
languages, makes mundane general-purpose programming tasks difficult.
Second, building and distributing Fortran software has been relatively
difficult, especially for newcomers to the language.
Third, Fortran does not have a community maintained compiler like Python, Rust
or Julia has, that can be used for prototyping new features and is used by the
community as a basis for writing tools related to the language.
Finally, Fortran has not had a prominent dedicated website\textemdash
an essential element for new users to discover Fortran, learn about it,
and get help from other Fortran programmers.
In the same way, Fortran is no longer widely taught to university students 
or valued as a useful skill by industry.
As a consequence, adoption of new users has been stagnating,
large scientific Fortran projects have been migrating to other languages,
and the communities of Fortran programmers remained scattered and isolated.

To address these issues, a new open source Fortran community called
\textit{Fortran-lang} was formed in December 2019 \cite{curcic2021toward}.
Its initial conception came about in the J3 Fortran Proposals repository on
GitHub, described in a later section, where it became clear that there was a
need for modern tooling and an improved web presence for the Fortran community.
We describe these projects in more detail in the rest of this section.

\subsection{Online channels}
Modern computer languages generally have a website,
offering resources to new and existing users all in one place.
Several online resources have existed for some time\textemdash
notably, the \textit{comp.lang.fortran} Usenet group, the Fortran Wiki\footnote{\url{http://fortranwiki.org}}, and the \url{https://fortran90.org} site\textemdash
however, there has never been a central web presence run by the community until now.
The new \href{https://fortran-lang.org}{https://fortran-lang.org} website provides learning resources,
a list of open source and commercial compilers,
guides to contribute to the community projects,
a Fortran package index,
and a monthly newsletter.

Besides the main website, Fortran-lang manages a traditional mailing list for news and announcements,
a \href{https://twitter.com/fortranlang}{@fortranlang} Twitter account,
and the Fortran Discourse\footnote{\url{http://fortran-lang.discourse.group}} which provides a friendly and welcoming online discussion board for all Fortran programmers\textemdash
existing and aspiring alike.
The new Discourse forum has proved a popular success, totaling more than 300 users in one year, with threads covering a wide variety of interesting Fortran-related topics.

At the core of the new Fortran-lang community have been its flagship projects\textemdash
the Fortran Standard Library and the Fortran Package Manager\textemdash
which were identified early on as critical for improving the Fortran ecosystem and tooling.
The development of these projects has been performed openly and collaboratively thanks to the distributed version control software \textit{Git} and the popular hosting platform \textit{GitHub}.
GitHub allows everyone to clone an open source project, test it,
modify the code and offer it to the community via a ``Pull Request'' to be reviewed.
Everyone can therefore easily contribute to the projects of the Fortran-lang GitHub organization\footnote{\url{https://github.com/fortran-lang}} and,
at the time of writing, there are 164 contributors to the main Fortran-lang projects through code contributions and discussions.
This includes six student interns, funded by the Google Summer of Code initiative, to work on Fortran-lang projects in 2021.
In the following sections, we present the new standard library and package manager projects.

\section{THE FORTRAN STANDARD LIBRARY}

\subsection{Motivation}

Although the Fortran Standard specifies a number of built-in procedures and
modules (collectively called \emph{intrinsics}), it does not define a standard
library that is common to languages like C, C++, Python, or Rust. Currently, Fortran 2018
defines 168 intrinsic procedures and a small number of constants
and derived types, but no widely used data structures like linked
lists and dictionaries.

Thus, commonly used features\textemdash
string and error handling, high-level I/O, statistical methods\textemdash
are programmed by users in-house, creating many redundant implementations with
different levels of correctness and rigor.
Proliferation of tailor-made,
common functionalities is inefficient and error-prone, and
discourages the adoption of the language in emerging fields,
reducing its future relevance.
For example, Fortran has not seen much application in the computationally-intensive area of neural networks for machine
learning despite being a performant parallel array-oriented language.
Fortran neural-network frameworks do exist \cite{Curcic2019parallel},
but the majority of deep learning is performed in Python.
Though not originally designed for high-performance numerical work,
Python has attracted machine learning practitioners
with its ecosystem of packages and rich standard library.
Furthermore, Fortran intrinsics have often been formally
standardized before being prototyped in compilers, thus slowing down the
modernization of the language and creating opportunities for poor API design.

The Fortran standard library (\textit{stdlib}) project aims at providing
community-developed, robust and reusable reference implementations of commonly
used procedures, in a versioned and well-tested library.
Consequently, \textit{stdlib} intends to reduce practitioners duplication of
effort and to help lower the bar for Fortran's adoption.
A future aim is to collaborate with the Fortran
Standard Committee and compiler developers to incorporate relevant procedures
into the language intrinsics and provide optimized implementations alongside compilers.

\subsection{Scope}

In addition to standardizing common interfaces for core functionalities\textemdash
like input and output (I/O), containers, and filesystem access \textit{etc.}\textemdash
the scope of \textit{stdlib} is akin to that of NumPy~\cite{numpy}, SciPy~\cite{scipy}, and MATLAB,
thus providing procedures useful for science, engineering, and mathematics.
As of July 2021, \textit{stdlib} contains 17 modules
that contain procedures for handling errors and optional arguments, working with bitsets,
facilitating I/O operations, linear algebra, logging, numerical integration, descriptive statistics, sorting, and strings.
More information can be found in full API documentation online (\href{https://stdlib.fortran-lang.org}{https://stdlib.fortran-lang.org}) generated by the FORD tool~\cite{MacMackin2018}.

\subsection{Development}

To provide generic interfaces for all supported types, kinds, and ranks,
\textit{stdlib} uses the Fypp\footnote{\url{https://github.com/aradi/fypp}} preprocessor to generate code from templates.
This is required since generic programming in Fortran is currently limited to mandate for every combination of type,
kind and rank an implementation.
However, all such procedures are then usually invoked using a common generic name.

To ensure cross-platform support, \textit{stdlib}
exploits GitHub's Continuous Integration (CI) pipeline
to test all procedures, derived types, and modules on the main branch
and on any changes proposed in pull requests.
The CI pipeline is run across a combination of different compilers and operating systems (Table \ref{tab:stdlib_compilers});
beyond this, \textit{stdlib} should be supported on any platform with Python (for Fypp) and a Fortran 2008 compliant compiler.

\begin{table*}[ht]
	\centering
	\caption{Operating systems and compilers on which the Fortran Standard Library is regularly tested}
	\label{tab:stdlib_compilers}
	\begin{tabular}{cccc}
	\hline
	Operating system & Architecture & Compiler & Version \\
	\hline
	\multirow{2}{*}{MacOS Catalina 10.15} & \multirow{2}{*}{x86\_64} & GCC Fortran & 9, 10, 11 \\
	&                          & Intel oneAPI classic & 2021.1  \\ \hline
	\multirow{2}{*}{Ubuntu 20.04} & \multirow{2}{*}{x86\_64} & GCC Fortran & 9, 10, 11  \\
	&                         & Intel oneAPI classic & 2021.1 \\ \hline
	\multirow{2}{*}{Microsoft Windows Server 2019} & \multirow{2}{*}{x86\_64} & GCC Fortran (MSYS) & 10  \\
	&             & GCC Fortran (MinGW) & 10 \\
	\hline
	\end{tabular}
\end{table*}

\section{THE FORTRAN PACKAGE MANAGER}

\subsection{Motivation}
The translation of a compiled language source code to machine instructions
in a large project, involves scheduling source files for compilation, taking
into account any inter-file dependencies, and linking the resulting objects
together and with any third-party dependencies.
A typical Fortran project relies on hand-coded Makefiles (not portable), CMake
(complicated to learn and use), or custom build scripts (difficult to
maintain) to perform the build process.
Thus, a new Fortran programmer not only needs to learn Fortran and how to apply
it to solve their problem, but they also need to learn one or more build
systems.
Moreover, such language-agnostic build systems have no easy way to use existing Fortran
libraries as dependencies in new projects.
For experienced Fortran programmers, this has been the way of life\textemdash
but for newcomers it is a serious barrier to entry.

Newer languages like Rust (\href{https://doc.rust-lang.org/cargo}{cargo}),
Racket (\href{https://docs.racket-lang.org/pkg/getting-started.html}{raco}),
Haskell (\href{https://www.haskell.org/cabal/}{cabal}) and
Python (\href{https://www.python.org/dev/peps/pep-0518/}{distutils})
have simplified the process of fetching dependencies and building/running the ensemble by abstracting much of the aforementioned complexity through a language-specific build system or package manager.

The goal for the Fortran Package Manager (\textit{fpm}) is to have a Fortran-specific build system and package manager to simplify compiling Fortran code and using third-party dependencies.
\textit{Fpm} removes the need to learn about and maintain complex build systems thereby simplifying the learning curve for newcomers.

\begin{figure*}[t]
	\centering
	\includegraphics[width=0.75\textwidth]{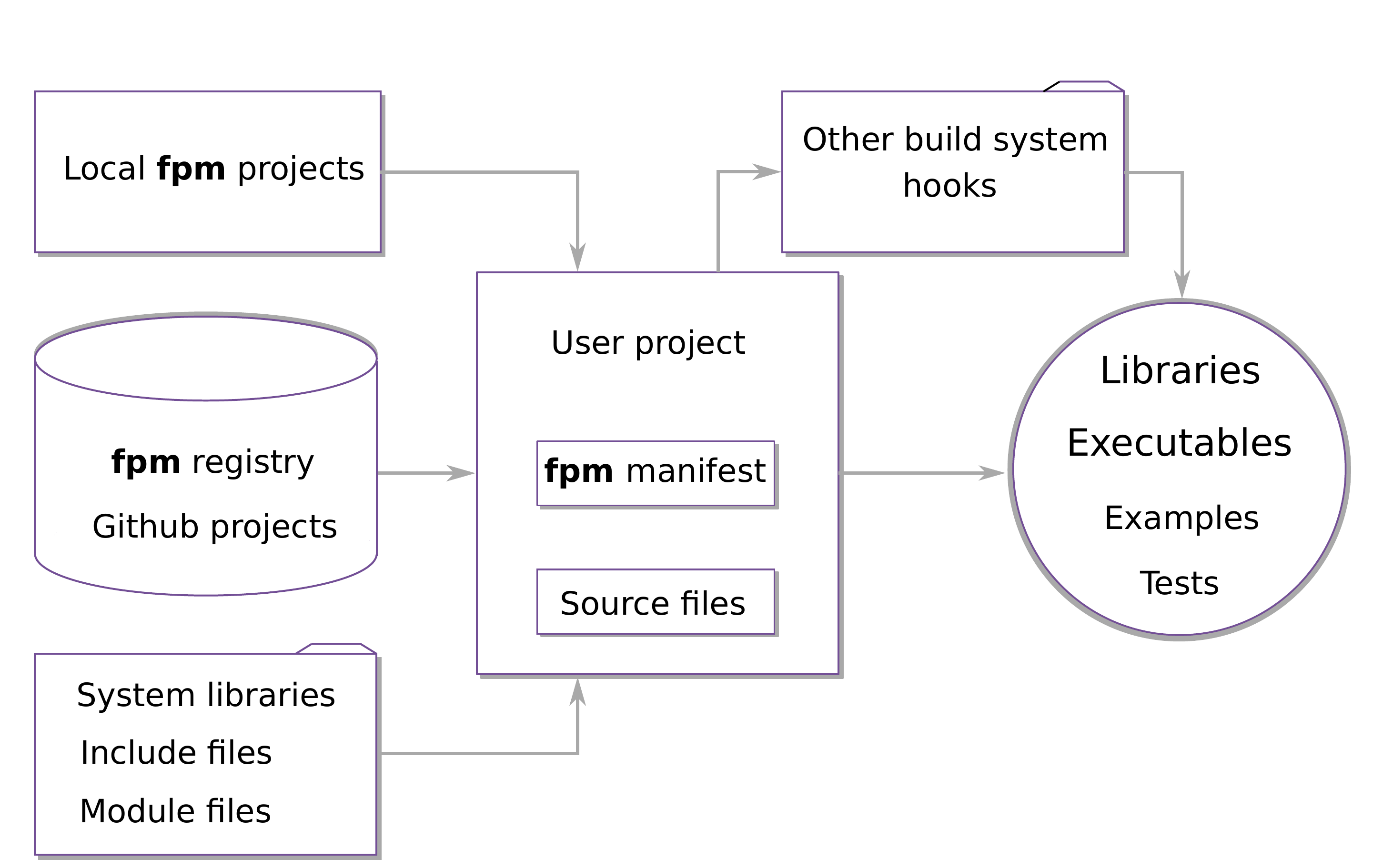}
	\caption{Fortran package manager workflow}
	\label{fig:fpm}
\end{figure*}

\subsection{Design Considerations}
\textit{Fpm}'s functionality is accessed by an intuitive command-line interface modeled after Rust's package manager Cargo.
It covers common Fortran project tasks including:
creating new projects; compiling and linking; running applications and tests; and fetching third-party dependencies to use within the current project (Figure \ref{fig:fpm}).

The minimal requirement for packages to be used with \textit{fpm} is the inclusion of the package manifest file,
written using the TOML\footnote{\url{https://toml.io}} format.
It allows \textit{fpm} to automatically find sources, executables, and unit tests,
if they are placed within the default folder structure.
The manifest file can contain additional metadata (author, maintainer, license, copyright, \textit{etc.}) or options for finer control of \textit{fpm}'s behavior.
\textit{Fpm} is written in Fortran and built by itself or from a single source file with just a Fortran compiler.

A key feature of \textit{fpm} is the ability to easily reuse code between projects.
Reusing compiled Fortran modules has been difficult due to missing Application Binary Interface~(ABI) compatibility between compilers, and even between versions of the same compiler.
\textit{Fpm} solves this by providing first-class support for handling dependencies between Fortran projects while keeping builds with different compilers and options separate.
Additionally, complex compiler configuration is abstracted away from the user while retaining access to commonly used debugging options for development and recommended optimization settings for production.

\textit{Fpm}'s first class support for dependency management makes
it straightforward to specify dependencies and control their exact version.
If a project and all of its dependencies are handled by \textit{fpm}, stronger guarantees are possible for effortless and reproducible builds.
Other aspects that can be controlled are the compiler, the compiler version, the build flags, and the operating system.
With a project's dependencies and compiler flags fully specified, one only needs to define an operating system, compiler and its version to reproduce a specific build, which is manageable even for new users.
This solves what has historically been an intractable problem for many Fortran programmers and their projects.

\subsection{Future features}
Ongoing work is currently focused on providing fine-grain control of compiler flags under different build \textit{profiles} within the manifest file.
This process is complicated by the many different available Fortran compilers (each with their own flags), the intricate semantics that are associated with certain flags, and linking with platform-specific resources.
\textit{Fpm}'s end goal in this respect is to abstract away all of the complications in a compiler-agnostic manner.
Furthermore, \textit{fpm} will eventually provide native support for using features such as \textit{MPI}, \textit{OpenMP}, \textit{OpenACC}, and \textit{Coarrays}.

To help newcomers who are unfamiliar with command-line interfaces, a cross-platform graphical user interface is in development. Alongside common tasks such as building and installing, the prototype allows setting build options, such as ``debug'' or ``release'' as well as browsing the online \textit{fpm} registry and navigating to the webpage of a selected package.

Finally, peripheral features common to other language-specific package managers are planned, including the ability to query and download packages from one or more central registries with the command line, and integrated support for documentation generation.

\section{THE FUTURE IS BRIGHT}

In addition to the \textit{stdlib} and \textit{fpm} projects, there are several other ongoing community efforts to improve the Fortran language, its compilers, and its outreach.

\subsection{Evolving the language as a community}

Development of the Fortran language standard follows the ISO standardization
workflow which is led by national Fortran standards committees and the working
group WG5 of the ISO/IEC\,JTC\,1/SC22 standardization subcommittee.  Committee
members are typically delegated by participating organizations, such as national
labs, research institutes, or compiler and hardware vendors.  While the formal
standardization procedure is beneficial for maintaining quality and robustness
of the language, the period for a proposed feature to become part of the
standard is consequently very long as compared to popular non ISO-standardized
languages like Python or Rust.  Moreover, the language development is quite
decoupled from the wider community of Fortran programmers.

To address this shortcoming, two members of the U.S. Fortran Standards Committee suggested establishing a public Git repository\footnote{\url{https://github.com/j3-fortran/fortran_proposals}} for language proposals.
The repository was created in October 2019 on GitHub
and became an immediate success, highlighting a long-standing desire within the community to participate in developing the language.
Community members are now able to easily propose and discuss additions to the language which, after discussion and consensus, can be made into formal proposals for discussion by the committee.
More than 190 issues and 25 pull requests have been created so far, and some of the suggested features are expected to become part of the Fortran future 202Y standard.
Similarly, the highly discussed topic of improving generic programming in Fortran has had a separate working group and public repository\footnote{\url{https://github.com/j3-fortran/generics}} set up to address this widely sought after functionality.

\subsection{Compiler development}

\begin{table*}[ht]
		\centering
		\caption{Summary of free-to-use multi-platform Fortran compilers}
		\label{tab:fortran_compilers}
	\begin{tabular}{rll}
		\hline
		\textbf{gfortran}              & Open source, GPLv3        & Full support for F2003, partial support for F2008 and F2018 \\
		\textbf{Classic flang}         & Open source, Apache-2.0   & Full support for F2003, to be superseded by LLVM flang      \\
		\textbf{LLVM flang}            & Open source, Apache-2.0   & Under development, full support for parsing F2018           \\
		\textbf{LFortran}              & Open source, BSD-3-Clause & Under development, full support for parsing F2018           \\
		\textbf{Intel Classic (ifort)} & Proprietary, Intel  & Full support for F2018                                      \\
		\textbf{Intel LLVM (ifx)}      & Proprietary, Intel  & Beta development, full support for F95                                                \\
		\textbf{nvfortran}             & Proprietary, NVIDIA & Full support for F2003, partial support for F2008          \\
		\hline
	\end{tabular}
\end{table*}

Fortran has several free-to-use compilers (Table \ref{tab:fortran_compilers}), of which
GFortran\textemdash{}part of GCC\textemdash
is the most advanced open source option (licensed under GPL).
It supports most of the latest Fortran standards and is the recommended option
for a production compiler today.

Three new Fortran compilers, based on LLVM, are currently under development: Flang,
which is part of the LLVM project, ifx by Intel and LFortran\footnote{\url{https://lfortran.org}}.

LFortran is a modern open-source (BSD licensed) interactive Fortran compiler
built on top of LLVM.
Its goal is to become a community supported compiler
that can be used to prototype new features and develop new tools for Fortran.
It can execute user’s code interactively for
exploratory work at the command line or in a Jupyter notebook (much like Python,
MATLAB or Julia) as well as compiling to binaries.
LFortran is currently under active development with the aims to support all of
Fortran 2018 and target modern architectures such as multi-core CPUs and GPUs.

LFortran has separate Abstract Syntax Tree (AST) and Abstract Semantic Representation
(ASR) structures which allows for greater flexibility in the compiler.
For example, the AST representation can be used as an automatic code formatter.
Likewise, the standalone ASR can be the input of multiple backends for code generation.
Indeed, LFortran has an LLVM backend, a fast x86 one, and a code generator
for C++.

\subsection{Research and dissemination}

A great landmark for Fortran was the inaugural FortranCon
2020\footnote{\url{https://tcevents.chem.uzh.ch/event/12/}}, an international conference which took place on July
2\textendash{}4, 2020, in Z{\"u}rich, Switzerland.  Lead by Tiziano M{\"u}ller
(Department of Chemistry, University of Z{\"u}rich) and Alfio Lazzaro (Hewlett
Packard Enterprise, Switzerland), the conference attracted more than 270
participants for two full days of talks and a half-day workshop on
object-oriented programming in Fortran.
The keynote
talk \textit{``Fortran 2018\,…\,and Beyond''} was delivered by Steve Lionel giving an overview of the recent standardization
efforts; this was followed by talks covering community projects, scientific applications,
compilers, programming tools, parallel programming, and
interoperability with other languages.
FortranCon 2021 was held in September 2021, with similar scope of presentation and attendance as the inaugural one.
We hope that future FortranCon events continue to amplify the work of Fortran
practitioners, enhance community-building, and increase language adoption.

\section{CONCLUSIONS}

The Fortran language has evolved enormously since its invention over sixty years ago such that it continues to offer a wealth of useful high-level abstractions for engineers and scientists.
Moreover, the language still retains its high performance through its array oriented design, strong static guarantees, and its native support for shared memory and distributed memory parallelism.

In this article we have exposed the lesser-known, high-level features of Fortran and highlighted the key advantages for code longevity and stability.
Similarly, we have identified the main limitations to using Fortran in the modern age, and have described recent efforts by the new Fortran-lang online community at addressing these.
In particular are the provision of a central online web presence with a moderated discussion board; the collaborative development of a Fortran standard library and Fortran package manager; and community led efforts to develop the language and its compilers.

With a burgeoning ecosystem of tooling and open source code, a growing online community, and ongoing development of the language and its compilers, Fortran is on track for continued and increased growth into the future.

\section{ACKNOWLEDGMENT}

We gratefully acknowledge all contributors to the Fortran-lang open source projects, Fortran Discourse participants, compiler developers, and Fortran Standard Committees.
We thank Pierre De Buyl and two anonymous reviewers whose comments helped improve this article.

\bibliographystyle{IEEEtran}
\bibliography{IEEEabrv,bibliography}

\begin{IEEEbiography}{Laurence J. Kedward}{\,} is a research associate in the department of aerospace engineering at the University of Bristol, UK where his research interests are computational fluid dynamics and numerical optimization.
\end{IEEEbiography}

\begin{IEEEbiography}{B\'alint Aradi}{\,}is a senior researcher at the Bremen Center of Computational Materials Science at the University of Bremen, Germany where he develops and applies quantum mechanical simulation methods to explore materials properties on the atomistic scale.
\end{IEEEbiography}

\begin{IEEEbiography}{Ond\v{r}ej \v{C}ert\'{i}k}{\,}is a scientist at Los Alamos National Laboratory. He is the original author of the SymPy, SymEngine, and LFortran open source projects, and co-founder of Fortran-lang.
\end{IEEEbiography}

\begin{IEEEbiography}{Milan Curcic}{\,}is an assistant scientist at the
University of Miami. He specializes in ocean wave physics and numerical
weather prediction.
\end{IEEEbiography}

\begin{IEEEbiography}{Sebastian Ehlert}{\,}is a PhD student at the University of Bonn (Germany),
in the Mulliken Center for Theoretical Chemistry,
where he is working on the development of fast and robust semiempirical electronic structure methods.
\end{IEEEbiography}

\begin{IEEEbiography}{Philipp Engel}{\,}is a PhD student in Geodesy at the
    Technical University of Berlin (Germany), where his research includes sensor
    networks, deformation monitoring, and spatial data analysis.
\end{IEEEbiography}

\begin{IEEEbiography}{Rohit Goswami}{\,} is a Rannis funded doctoral researcher
in the Jonsson group at the Science Institute and Faculty of Physical Sciences
at the University of Iceland where his work revolves around scientific high
performance computing and Bayesian statistical learning for scalable ab-initio
studies of magnetic materials. He also maintains f2py within NumPy as a software
engineer at Quansight Labs.
\end{IEEEbiography}

\begin{IEEEbiography}{Michael Hirsch}{\,}is a research scientist with the Center for Space Physics at Boston University. He uses Fortran to model sensor and system performance from subsurface to space including the atmospheric and ionospheric response.
\end{IEEEbiography}

\begin{IEEEbiography}{Asdrubal Lozada-Blanco}{\,}is a postdoctoral researcher in the S\~{a}o Carlos Institute of Physics
at the University of S\~{a}o Paulo (Brazil), where he develops simulation methods in statistical mechanics.
\end{IEEEbiography}

\begin{IEEEbiography}{Vincent Magnin}{\,}is an associate professor at the University of Lille (France),
in the Institut d’Electronique, de Micro\'{e}lectronique et de Nanotechnologie,
where his interests are the numerical modeling and optimization of optoelectronics and photonics devices.
\end{IEEEbiography}

\begin{IEEEbiography}{Arjen Markus}{\,}is a senior consultant at the Deltares research institute, where he is responsible for several numerical modelling programs, specifically related to water quality. He has contributed to the ACM Fortran Forum and in 2012 he published the book "Modern Fortran in practice".
\end{IEEEbiography}

\begin{IEEEbiography}{Emanuele Pagone}{\,}is Research Fellow in Sustainable Manufacturing Modelling at Cranfield University where he simulates energy and manufacturing systems to support environmentally sustainable multi-disciplinary decisions.
\end{IEEEbiography}

\begin{IEEEbiography}{Ivan Pribec}{\,}is a PhD student in the TUM School of Life Sciences at the Technical University of Munich, where he develops and applies lattice Boltzmann methods to explore and optimize chemical engineering unit operations.
\end{IEEEbiography}

\begin{IEEEbiography}{Brad Richardson}{\,}is a consultant and scientific software developer at Archaeologic, Inc.
His efforts focus on developer productivity through the development and use of better tools,
and the discovery and application of better software design principles.
\end{IEEEbiography}

\begin{IEEEbiography}{Harris Snyder}{\,}is a scientific computing engineer at Structura Biotechnology, where he works on data processing software for cryo-electron microscopy.
\end{IEEEbiography}

\begin{IEEEbiography}{John Urban}{\,} currently provides HPC consultation and support at the Bettis Atomic Power Laboratory,
for the design and development of nuclear power for the U.S. Navy.
His career has ranged from Engineering commercial power at Westinghouse to work at Cray, Digital, Compaq, HP, and SGI.
\end{IEEEbiography}

\begin{IEEEbiography}{J\'{e}r\'{e}mie Vandenplas}{\,}is a researcher in Animal
Breeding and Genomics at Wageningen University \& Research, where his interests are
quantitative genetics and genomic prediction related to livestock.
\end{IEEEbiography}

\end{document}